# Thermal Conductivity Anisotropy in Superconducting $UPt_3$

A. Fledderjohann and P.J. Hirschfeld

*Dept. of Physics, Univ. of Florida, Gainesville, FL 32611, USA.*

Recent thermal conductivity measurements on $UPt_3$ single crystals by Lussier et al. indicate the existence of a strong b–c anisotropy in the superconducting state. We calculate the thermal conductivity in various unconventional candidate states appropriate for the $UPt_3$ "B phase" and compare with experiment, specifically the $E_{2u}$ and $E_{1g}$ $(1, i)$ states predicted in some Ginzburg-Landau analyses of the phase diagram. For the simplest realizations of these states over spherical or ellipsoidal Fermi surfaces, the normalized $E_{2u}$ conductivity is found, surprisingly, to be completely isotropic. We discuss the effects of inelastic scattering and realistic Fermi surface anisotropy, and deduce constraints on the symmetry class of the $UPt_3$ ground state.

The heavy fermion superconductor $UPt_3$ is now generally thought to be "unconventional", in the sense that the order parameter exhibits a lower symmetry than that of the Fermi surface. The principal evidence for this conclusion has been the observation of a nontrivial phase diagram for the system in applied magnetic field and pressure. Earlier hints of unconventional behavior included transport and thermodynamic properties reported to vary as power laws in the temperature for $T \ll T_c$, as well as strong, temperature-dependent anisotropy in superconducting transport properties. [1]

Current Ginzburg-Landau (GL) theories of the $UPt_3$ phase diagram attempt to explain the multiplicity of superconducting phases observed by assuming an order parameter corresponding to 1) a higher dimensional irreducible representation of the normal state symmetry group, weakly split by a symmetry–breaking field; or 2) a mixture of two accidentally nearly degenerate representations. In most scenarios, the quartic terms in the GL free energy are chosen to reproduce, if possible, the expected nature of the low-temperature, low-field $UPt_3$ "B-phase", characterized by a line of gap nodes in the hexagonal basal plane, and possibly point nodes along the c-axis. The prejudice in favor of this structure arises from early tranport experiments. [1,2] Qualitatively speaking, these experiments support a picture in which there is, for $T \ll T_c$, a higher density of excited quasiparticles with wave vectors in the basal plane. Attempts to measure this anisotropy in other experiments have not been uniformly successful, however. For example, measurements of the London penetration depth $\lambda_L$ in geometries chosen to maximize currents along chosen directions found $\Delta \lambda_L(T) \sim T^2$ with coefficients depending only weakly on direction. [3] A $\mu SR$ experiment reporting considerable anisotropy in the same quantity [4] may in fact have been dominated by extraneous effects. [5] Explicit evidence in favor of strong gap anisotropy in the superconducting state involving measurements in different directions in the same sample is in fact rather limited. [1,6] The recent thermal conductivity measurements of Lussier et al. are therefore of considerable interest, both as hard qualitative evidence for anisotropy and for the opportunity to make quantitative comparisons with theory.

The purpose of this paper is to apply the theory of heat conduction in a weak-coupling BCS superconductor [7] to the data of Lussier et al., [8] assuming unconventional candidate order parameters of various types currently proposed as ground states for the $UPt_3$ system. While it has been argued that fits to transport coefficients at low temperatures are insufficient to fix the order parameter symmetry, given uncertainties in the modelling of the scattering amplitude, anisotropy ratios of transport quantities should depend much more weakly on these details, thereby reflecting more directly the actual anisotropy of the superfluid condensate. We therefore focus primarily in what follows on the ratio $\kappa_c/\kappa_b$ between the conductivities measured for heat currents directed along the c– and b–axes, respectively.

*Normal state.* A good deal of important information regarding normal state scattering parameters was extracted by Lussier et al. from their measurements above the critical temperature $T_c = 0.5K$. In this region their data could be well fit by the form $\kappa_{N,i}^{-1}(T)/T = (a_i + b_i T^2)$, where $i = b, c$ indexes the eigenvalues of the thermal conductivity tensor. The authors argue convincingly that i) the heat conductivity they measure is almost entirely electronic in origin, noting that a substantial phonon contribution would not allow for the observed isotropy of the Lorenz number $\kappa \rho/T$. Furthermore, ii) both anisotropy ratios $a_b/a_c$ and $b_b/b_c$ are empirically found to be approximately 2.7, suggesting strongly that the anisotropy in the electronic thermal conductivity $\kappa_N = c_v v_F^2 \tau/3$ arises from anisotropy in the Fermi velocity rather than in the relaxation time $\tau$ (Here $c_v$ is the electronic specific heat and $v_F$ the Fermi velocity). Since $v_F \sim m^{-1}$, this is consistent with the mass ratio $m_b/m_c = 1.7$ deduced from upper critical field measurements. [9] We therefore assume i) that heat conduction occurs entirely through electronic transport, and ii) restrict our treatment of collisions to s-wave scattering.

Impurity scattering is treated within a self-consistent t-matrix approximation [10,11] for the self-energy, which in the normal state reduces to an s-wave scattering rate $\Gamma$, while for inelastic processes we simply broaden all quasiparticle states by a phenomenological inverse lifetime $\tau_{in}^{-1} = \alpha T^2$. This crude model will suffice for our purpose of fixing the relative importance of inelastic contributions to superconducting state transport. The coefficient $\alpha$ may be determined from a fit to the Lussier et al. normal state coefficient $b$ to satisfy $0.5\alpha \simeq \Gamma/T_c^2$; thus at the critical temperature the inelastic and elastic scattering rates are roughly comparable. Thus at least for the temperature range close to $T_c$, inclusion of inelastic processes is essential if a quantitative comparison with experiment is to be attempted. The measured residual linear term $a$ in $\kappa_N^{-1}(T)$ now allows for a crude determination of the magnitude of the impurity scattering rate. Using the usual kinetic theory result for the thermal conductivity, $\kappa_N = c_v(T_c)v_F^2\tau/3$, and using the linear specific heat coefficient $\gamma = 0.42 J/K^2 \cdot$ mole and the de Haas–van Alphen measurements of Taillefer and Lonzarich, [12], we extract $\Gamma/T_c \simeq 0.2 \pm 0.1$. This is substantially larger than impurity scattering rates which were earlier considered typical of nominally pure samples, [10,13,14] but previous estimates as to the size of this quantity were not been based on reliable normal state information.

*Superconducting order parameter.* As discussed above, there is some evidence for an order parameter corresponding to the low-temperature "B-phase" of $UPt_3$ whose orbital structure manifests line nodes in the hexagonal basal plane. We therefore focus our attention on two such states, the $E_{1g}$ $(1,i)$ state proposed by various authors [10,15] and the $E_{2u}$ $(1,i)$ state with spin axis $\hat{d} \parallel \hat{c}$ discussed more recently by Norman [16] and Sauls [17]. Although the former is an even-parity ("d-wave") spin singlet state, and the latter an odd-parity ("f-wave") spin triplet state, the spin structure of the two order parameters is irrelevant for the calculation of the thermal conductivity, and will be neglected here. The orbital parts of the order parameters $\Delta_\mathbf{k}$ then have symmetry $\hat{k}_z(\hat{k}_x + i\hat{k}_y)^n$, where $n = 1(2)$ for the $E_{1g}(E_{2u})$ state. Both states have a line of nodes in the basal plane and point nodes along the c-axis but differ significantly in the behavior of the order parameter in the vicinity of the point nodes. This difference, usually ignored, can have important consequences for the anisotropy of transport coefficients, as we show below. In the 1D (orbital) representation put forward by Machida et al. [18], the most likely states are the $A_{2u}$ state, which supports quasiparticle excitations with spectrum identical to the $E_{2u}$ state given above, or the polar state with $\Delta_k \propto \hat{k}_z$. In what follows we will therefore refer to $E_{1g}$, $E_{2u}$ or polar, meaning however the broader classes of states with the orbital structures given above.

The exact form of the order parameter will depend on the detailed form of the Fermi velocity $\mathbf{v}_{F\mathbf{k}} = \partial \epsilon(k)/\partial \mathbf{k}|_{\epsilon_F}$ and angle-resolved density of states, as well as the distribution of the pair potential weight over the various Fermi surface sheets. For the moment we ignore all such complications, and assume that the candidate order parameters in question are represented over a spherical Fermi surface by $\Delta_\mathbf{k} = \Delta_0(T) \cdot \sin 2\Theta \, e^{i\phi}$ for the $E_{1g}$ case and $\Delta_\mathbf{k} = \frac{3\sqrt{3}}{2}\Delta_0(T) \cdot \sin^2\Theta \cos\Theta \, e^{2i\phi}$ for the $E_{2u}$ case, where $\Delta_0(T) \equiv \max\{|\Delta_\mathbf{k}|\}_{FS}$.

*Thermal Conductivity.* The thermal conductivity $\kappa$ is evaluated using a Kubo formula for the heat-current response similar to the original treatment for an s-wave superconductor. [7] The generalization to unconventional states has been discussed previously by several groups. [10,11,19,13] In the limit of vanishing impurity concentrations, identical results were also obtained by Arfi et al. [14] using a transport equation method.

The Kubo formula approach begins with an impurity-averaged single-particle propagator which in a Nambu matrix notation reads

$$\underline{g}(\mathbf{k},\omega) = \frac{\tilde{\omega}\underline{\tau}^0 + \xi_\mathbf{k}\underline{\tau}^3 + \underline{\Delta}_\mathbf{k}}{\tilde{\omega}^2 - \xi_\mathbf{k}^2 - |\Delta_\mathbf{k}|^2} \quad (1)$$

where $\underline{\tau}^i$ represent the Pauli matrices. Here, we have already exploited the assumed particle-hole symmetry of the normal state, as well the symmetries of the gap functions which lead to vanishing off-diagonal scattering self-energy contributions. In this limit, only self-energy contibutions to the frequency $\omega$, namely $\tilde{\omega} = \omega - \Sigma_0$ need to be included. [13] The self-energy $\Sigma_0$ due to the elastic impurity scattering is treated in a self-consistent $T$-matrix approximation and is given by $\Sigma_0 = \Gamma G_0/(c^2 - G_0^2)$, where $\Gamma = n_i n/(\pi N_0)$ is the unitarity limit scattering rate depending on the concentration of defects $n_i$, the electron density $n$, and the density of states at the Fermi level $N_0$. The quantity $c \equiv \cot\delta_0$ parameterizes the scattering strength of an individual impurity through the s-wave phase shift $\delta_0$. In this work we consider only unitarity limit scattering $c = 0$ since it is clear that weak scattering will lead to a weak temperature dependence inconsistent with experiment for the states in question. [20,21] The integrated propagator is $G_0 = (1/2\pi N_0)\sum_\mathbf{k}\text{Tr}\{\underline{\tau}^0 \underline{g}(\mathbf{k},\omega)\}$. The equation for the self-energies are now solved self-consistently together with the gap equation $\Delta_\mathbf{k} = -T\sum_{\omega_n}\sum_{\mathbf{k}'}V_{\mathbf{kk}'}(1/2)\text{Tr}\{\underline{\tau}^1 \underline{g}(\mathbf{k}',\omega_n)\}$.

The bare heat current response is given by a convolution of the Green's function $g$ with itself at zero external frequency and wave vector weighted with the bare heat current vertex $\omega\mathbf{v}_{F\mathbf{k}}\underline{\tau}^3$. [7] Impurity scattering vertex corrections to current-current correlation functions have been shown vanish identically for even parity states ($\Delta_\mathbf{k} = \Delta_{-\mathbf{k}}$). [13] Even for odd parity states, such corrections vanish in the unitarity limit considered here. For the diagonal thermal conductivity tensor one obtains

$$\frac{\kappa_i(T)/T}{\kappa_{N,i}(T_c)/T_c} = \frac{9}{\pi^2}\Gamma\int_0^\infty d\omega\left(\frac{\omega}{T}\right)^2\left(\frac{-\partial f}{\partial \omega}\right)K_i(\omega,T) \quad (2)$$

$$K_i(\omega, T) = \frac{1}{\tilde{\omega}'\tilde{\omega}''} \left\langle \hat{\mathbf{k}}_i^2 \cdot \frac{\tilde{\omega}^2 + |\tilde{\omega}|^2 - 2|\Delta_{\mathbf{k}}|^2}{\sqrt{\tilde{\omega}^2 - |\Delta_{\mathbf{k}}|^2}} \right\rangle_{\hat{\mathbf{k}}} \quad (3)$$

where $\tilde{\omega}'$ and $\tilde{\omega}''$ are the real and imaginary parts of $\tilde{\omega}$ and $f$ is the Fermi function.

We have evaluated Eqs(2-3) numerically for the two candidate states, using impurity scattering rates consistent with the estimate given above, and compared with the data of Lussier et al. [8] If, as we have argued, all normal state anisotropy is reflected in the factors of $v_F^2$ for each direction $b, c$, it is simple to extract the anisotropy introduced by pair correlations by plotting the *normalized* thermal conductivity $(\kappa(T)/T)/(\kappa_N(T_c)/T_c)$, as in Figure 1. We have normalized all quantities to $T_{c-}$ since the amplitude corresponding to the $(1,i)$ states we consider first becomes finite there; the comparison with data should not be taken seriously in the temperature range close to $T_{c-}$, since in a proper theory involving the symmetry breaking field a small admixture of the A-phase $(1,0)$ state should be present below the transition.

We note several interesting aspects of these results. Firstly, the $E_{2u}$ state is found to yield an isotropic thermal conductivity within this approximation, $\kappa_c(T)/\kappa_b(T) = 1$. This is a consequence of the large number of thermally excited quasiparticles in the neighborhood of the quadratic polar point nodes, which make a contribution to the thermal current in both directions equal to that of the basal line nodes. The result may be shown analytically to hold exactly for all temperatures in the superconducting state, for any current-current correlation function in spherical symmetry. The $E_{2u}$ state is therefore clearly not compatible with the data of Lussier et al. in the current approximation (see, however, the discussion below).

The $E_{1g}$ state does exhibit roughly the proper anisotropy, as pointed out by Lussier et al., although the absolute magnitudes of the calculated thermal conductivities are too small by roughly 30% at intermediate temperatures in this approximation. We note finally that at low temperatures $T \lesssim (\Gamma T_c)^{1/2}$ a residual linear term $\kappa \sim a_0 T$ should be observable, at least in the b direction. [10,11] This "gapless" behavior reflects a corresponding residual density of quasiparticle states in the self-consistent treatment of impurity scattering for states with line nodes, absent in the theory of Arfi et al. [14]

*Inelastic scattering.* The discrepancy between the magnitude of the theoretical and experimental results in the temperature range near $T_c$, as well as the extracted magnitudes of the normal state relaxation components, suggests that some treatment of inelastic scattering is necessary if a semiquantitative fit is to be attempted. In the absence of a complete microscopic description of the electron-electron collision processes in this system, we simply assume that below the superconducting transition, the number of quasiparticles available for scattering varies as $T/\Delta_0$ due to the line nodes in the gap. [22] The model $\tau_{in}^{-1}$ is therefore a simple interpolation between the two limits

$$\tau_{in}^{-1}(T) \approx \begin{cases} \alpha T^2 \cdot T/\Delta_0(0) & T \ll T_c \\ \alpha T^2 & T \lesssim T_c \end{cases} \quad (4)$$

We find that our results do not depend significantly on the choice of interpolation or the prefactor of $T/\Delta_0$.

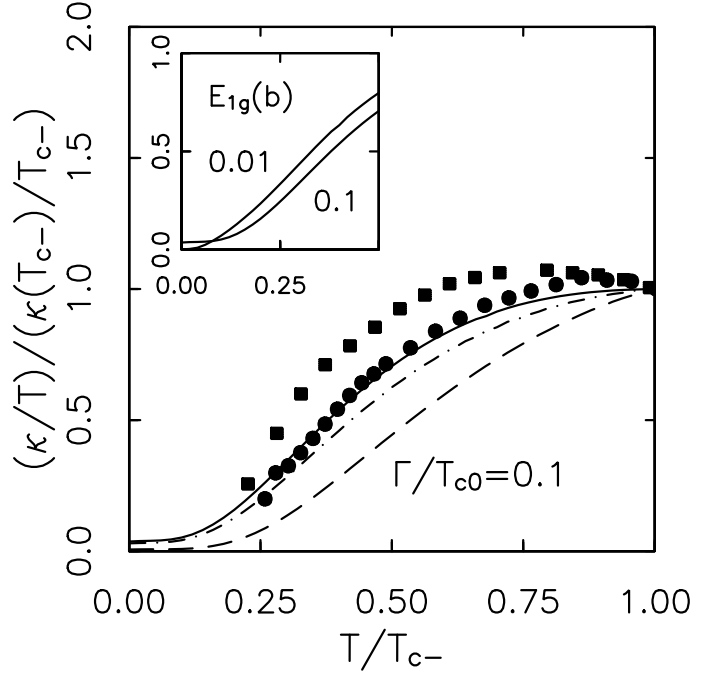

FIG. 1. Normalized thermal conductivity for impurity scattering rate $\Gamma/T_{c0} = 0.1$, no inelastic scattering. $T_{c0}$ is transition temperature in absence of impurities. Solid line (dashed lines): normalized $\kappa_b/T$ ($\kappa_c/T$) for $E_{1g}$ $(1,i)$ state. Dashed-dotted line: normalized $\kappa_b/T$ and $\kappa_c/T$ for $E_{2u}$ state. Squares (circles): $\kappa_b/T(\kappa_c/T)$ from reference [8] Insert: variation with $\Gamma$.

In order to get a qualitative picture about the influence of inelastic scattering contributions we now use a self-energy $\Sigma_0 = -\Gamma/G_0 - (i/2)\tau_{in}^{-1}(T)$, where $G_0$ is calculated as before but using $\tilde{\omega} = \omega - \Sigma_0$ using the modified $\Sigma_0$. In doing so we neglect not only the real part of the self-energy but also vertex corrections due to inelastic processes.

In Figure 2, we see that qualitatively, the effect of inelastic scattering is to increase the rise of the normalized $\kappa/T$ just below $T_c$ as expected. Quantitatively, it is intriguing to note that the inclusion of inelastic processes at this level of approximation is sufficient to bring the theoretical curves into rough agreement with experiment in the intermediate to high temperature range.

*Anisotropy.* All the above results depend directly upon our assumptions regarding the relaxation time $\tau$. The ratio of the thermal conductivities $\kappa_c/\kappa_b$ in the normal state is consistent with a model in which all anisotropy arises through the Fermi velocities. If this is true in the superconducting state as well, the anisotropy ratio is only very weakly dependent on the relaxation time, and thus constitutes an excellent probe of the intrinsic anisotropy of the superconducting condensate. In Figure 3 we plot the anisotropy ratio for the data of Lussier et al. and compare with our two theoretical candidate states. The figure suggests that the effective order parameter structure corresponds to a distribution of nodes intermediate between the $E_{1g}$ and $E_{2u}$ model states considered.

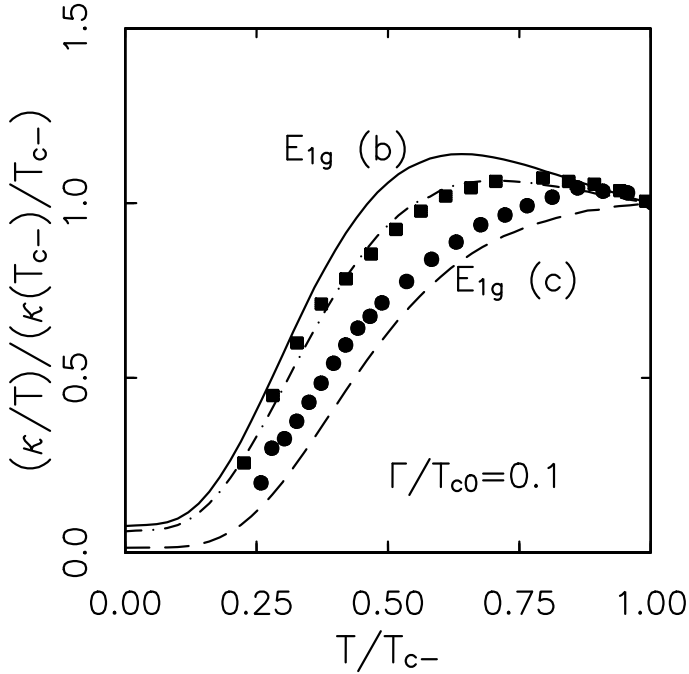

FIG. 2. Effect of inelastic scattering on thermal conductivity. Theory and experimental data as described in Fig. 1, but with inelastic self-energy effects included. Dashed-dotted line is for $E_{2u}$ state, both $\kappa_b$ and $\kappa_c$.

The most obvious source of error in the above analysis is the neglect of Fermi surface anisotropy, and at first glance it would appear that the simplest ellipsoidal model accounting for the mass anisotropy $m_b/m_c = 1.7$ observed in $UPt_3$ should be sufficient to break the artificial degeneracy between the two eigenvalues of $\kappa$ in the $E_{2u}$ state. A straightforward analysis shows, however, that this is not the case, and that all ellipsoidal anisotropy is reflected in the ratio $v_{Fc}^2/v_{Fb}^2$, which divides out of the normalized quantity plotted in Fig. 3. It is of course clear that inclusion of the complex structure of the true $UPt_3$ Fermi surface will give rise to some anisotropy, but significant effects of this kind appear unlikely. Finally, we note that the angular functions considered here are merely highly probable members of an infinite sequence of basis functions belonging to each irreducible representation. Inclusion of further such terms could also alter the anisotropy. While we can draw no definitive conclusions pending analysis of these effects, the data do appear to favor a B-phase ground state with line nodes in the basal plane and point nodes vanishing linearly near the poles.

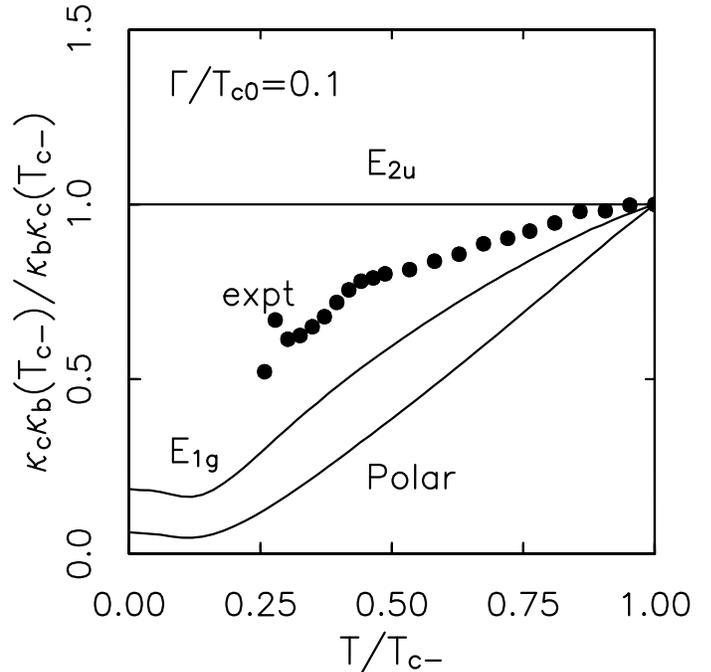

FIG. 3. Thermal conductivtity anisotropy $\kappa_c/\kappa_b$ of various states including inelastic scattering. Polar state ($\Delta_k = \Delta_0 \hat{k}_z$) is included for reference.

*Acknowledgements.* The authors gratefully acknowledge extensive discussions with L. Taillefer. AF was supported by the Deutsche Forschungsgemeinschaft.

4